\documentclass[aps,prl,twocolumn]{revtex4}

\usepackage{graphicx}
\usepackage{amsmath}

\begin{document}

\title{A microscopic description of the aging dynamics:
fluctuation-dissipation relations, effective temperature and
heterogeneities}

\author{A. Montanari}
\email{Andrea.Montanari@lpt.ens.fr}
\affiliation{CNRS-Laboratoire de Physique Th\'{e}orique de l'Ecole
Normale Sup\'{e}rieure, 24 rue Lhomond, 75231 Paris, France}
\author{F. Ricci-Tersenghi}
\email{Federico.Ricci@roma1.infn.it}
\affiliation{Dipartimento di Fisica, Universit\`a di Roma ``La
Sapienza'', Piazzale Aldo Moro 2, I-00185 Roma, Italy}

\date{\today}

\begin{abstract}
We consider the dynamics of a diluted mean-field spin glass model in
the aging regime.  The model presents a particularly rich
heterogeneous behavior.  In order to catch this behavior, we perform a
{\it spin-by-spin analysis} for a {\it given disorder realization}.
The results compare well with the outcome of a static calculation
which uses the ``survey propagation'' algorithm of M\'ezard, Parisi,
and Zecchina [Science{\it xpress} 10.1126/science.1073287 (2002)].  We
thus confirm the connection between statics and dynamics at the level
of single degrees of freedom.  Moreover, working with single-site
quantities, we can introduce a new response-vs-correlation plot, which
clearly shows how heterogeneous degrees of freedom undergo coherent
structural rearrangements.  Finally we discuss the general scenario
which emerges from our work and (possibly) applies to more realistic
glassy models.  Interestingly enough, some features of this scenario
can be understood recurring to thermometric considerations.
\end{abstract}

\maketitle

The description of the off-equilibrium dynamics in aging systems is
one of the major challenges in contemporary statistical mechanics.
Aging systems, like spin, structural and polymeric glasses
\cite{SITGES_ANGELL_STRUICK} are slowly evolving, heterogeneous
systems which do not reach thermal equilibrium, at low enough
temperatures, on any experimental time-scale.  In the last 20 years a
great effort has been devoted to study the dynamics of some
prototypical models, namely mean-field spin glasses
\cite{DynamicsReview}.  These models exhibit an extremely rich
behavior: slow relaxation, memory effects, aging.  Important tools
which have been introduced in this context are the {\em
off-equilibrium fluctuation-dissipation relation} (OFDR) \cite{OFDR},
and the {\em effective temperature} \cite{CUKUPE} one can derive from
that relation.  Such an effective temperature is, roughly speaking,
what would be measured by a thermometer responding on the time scale
on which the system ages.

One of the weak points of the results obtained so far is that they
focus on global quantities, e.g.\ correlation and response functions
averaged over the spins.  On the other hand, we expect one of the
peculiar features of glassy dynamics to be its {\em heterogeneity}
\cite{HETERO_EXP}.  For instance, correlation and response functions
of a particular spin depend upon its local environment
\cite{HETERO_SG}.  Two simple remarks are in order here: $(i)$ These
spin-to-spin fluctuations are non-zero even in the thermodynamic
limit; $(ii)$ They disappear when the average over quenched disorder
is taken.

Moreover the present definition \cite{CUKUPE} of effective temperature
has some problems.  Indeed it corresponds to what would be measured by
a specific, properly tuned, slow thermometer, while generalizations to
more generic thermometers give disagreeing results \cite{EXARTIER},
still to be clarified.

In this Letter, inspired by the new approach of~\cite{MEPAZE}, we
study the aging dynamics focusing on single-site correlation and
response functions.  In this way we are able to consider the
heterogeneities in the system and to define a microscopic, but {\em
site-independent}, effective temperature.

An interesting context for addressing these issues is provided by
diluted mean-field models.  In these models each spin interacts with a
finite number of other spins, just as in finite-dimensional models.
On the other hand, the absence of a finite-dimensional geometrical
structure makes them tractable from an analytic point of view.

We consider a ferromagnetic Ising model with 3-spin interactions,
defined by the Hamiltonian
\begin{equation}
\mathcal{H} = - \sum_{m=1}^{M} \sigma_{i_m} \sigma_{j_m} \sigma_{k_m}
\ ,
\end{equation}
where the $M$ triples $(i_m,j_m,k_m)$ are chosen randomly among the
$\binom{N}{3}$ possible ones.  Although ferromagnetic, this model is
thought to have a glassy behavior for $M > 0.818\, N$, due to {\em
self-induced} frustration \cite{RIWEZE}.

We work on a single sample with $M=N=100$, whose largest connected
component contains 96 sites.  We limited ourselves to such a small
sample because single-spin measures require {\em huge statistics}.

\begin{figure}[!ht]
\includegraphics[width=\columnwidth]{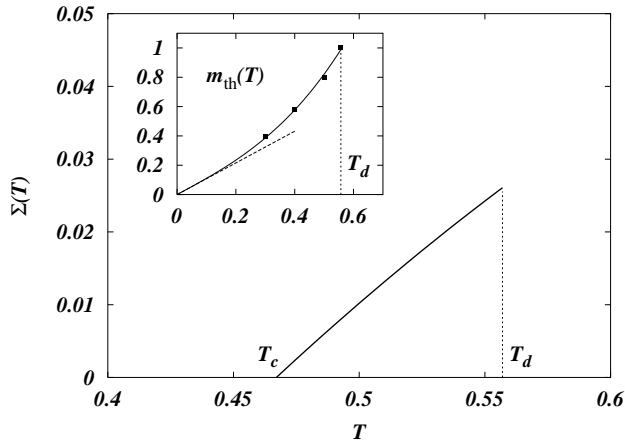}
\caption{The complexity $\Sigma$ and the 1RSB parameter for threshold
states $m_{\rm th}$ (Inset) versus the temperature for the sample
studied in this paper.  The continuous line in the inset is the
polynomial fit $m_{\rm th}(T) = 1.08\, T + 0.038\, T^2 + 2.17\, T^3$.}
\label{cplxT}
\end{figure}

We use the ``survey propagation'' (SP) algorithm of
Refs.~\cite{MEPAZE}, or better its generalization at finite
temperatures \cite{FDTSS2}, to compute the free energy density
$F(m,\beta)$ for our specific sample at one-step replica symmetry
breaking (1RSB) level.  In Fig.~\ref{cplxT} we report the complexity
$\Sigma(T) =\beta\partial_m F(m,\beta)|_{m=1}$.  The dynamic and
static temperatures are defined, respectively, as the points where a
non-trivial (1RSB) solution to the cavity equations first appears, and
where its complexity vanishes.  From the results of Fig.~\ref{cplxT}
we get the estimates $T_d = 0.557(2)$ and $T_c = 0.467(2)$.  In the
standard picture, the aging dynamics for discontinuous spin glasses is
dominated by ``threshold'' metastable states \cite{DynamicsReview}.
The corresponding 1RSB parameter $m_{\rm th}(T)$ can be computed by
imposing the condition $\partial_m^2[m F(m,\beta)]=0$.  We computed
$m_{\rm th}(T)$ on our sample for some temperatures below $T_d$, and
in the zero temperature limit, $m_{\rm th}(T) = \mu_{\rm th}T$, with
$\mu_{\rm th} =1.08(1)$. These results are summarized in the inset of
Fig.~\ref{cplxT}.

Another important outcome from the SP algorithm is the value of the
{\em local} Edwards-Anderson order parameter $q_{\rm EA}^{(i)}(m)$,
which depends on the 1RSB parameter $m$.  On threshold states, the
local order parameter is connected to the single-site correlation
function (defined below):
\begin{equation}
q_{\rm th}^{(i)} \equiv q_{\text{EA}}^{(i)}(m_{\rm th}) =
\lim_{t\to\infty} \lim_{t_w\to\infty} C_i(t_w+t,t_w) \ .
\end{equation}

We consider Metropolis dynamics starting from random initial
conditions, $\sigma_i(t\!=\!0)\!=\!\pm1$.  After a time $t_w$ we turn
on a small random magnetic field, $h_i = \pm h_0$, and we measure
single-spin correlation and integrated response functions
\begin{eqnarray}
C_i(t_w+t,t_w) & \equiv & \frac{1}{t} \sum_{t'=t}^{2t-1} \langle
\sigma_i(t_w+t') \sigma_i(t_w) \rangle \ , \\
\chi_i(t_w+t,t_w) & \equiv & \frac{1}{t h_0} \sum_{t'=t}^{2t-1}
\langle \sigma_i(t_w+t') \text{sign}(h_i) \rangle \, ,
\end{eqnarray}
where $\langle \cdot \rangle$ denotes the average over the Metropolis
trajectories and the random perturbing field.  The sums over $t'$ are
introduced in order to reduce the statistical error.  This
``experiment'' is repeat $N_{\text{runs}}$ times, each time with a
different thermal noise and perturbing field.  Typical values for
$N_{\text{runs}}$ range between $10^6$ and $5 \cdot 10^6$.

The first remark on the numerical data, is that the spins can be
clearly classified in two groups. Type-I spins behave as if the system
were in equilibrium: the corresponding correlation and response
functions satisfy time-translation invariance and the fluctuation
dissipation theorem (FDT).  Type-II spins are out of equilibrium
spins: their correlation and response functions are non-homogeneous on
long time scales and violate FDT.

Type-I includes isolated sites, but also 12 non-isolated sites.
Remarkably these sites are the ones for which the SP algorithm returns
$q^{(i)}_{\rm th} = 0$, i.e.\ they are paramagnetic from the static
point of view.  These sites can also be identified via a simple
algorithm \cite{LEAF_REM}.

\begin{figure}[!ht]
\includegraphics[width=\columnwidth]{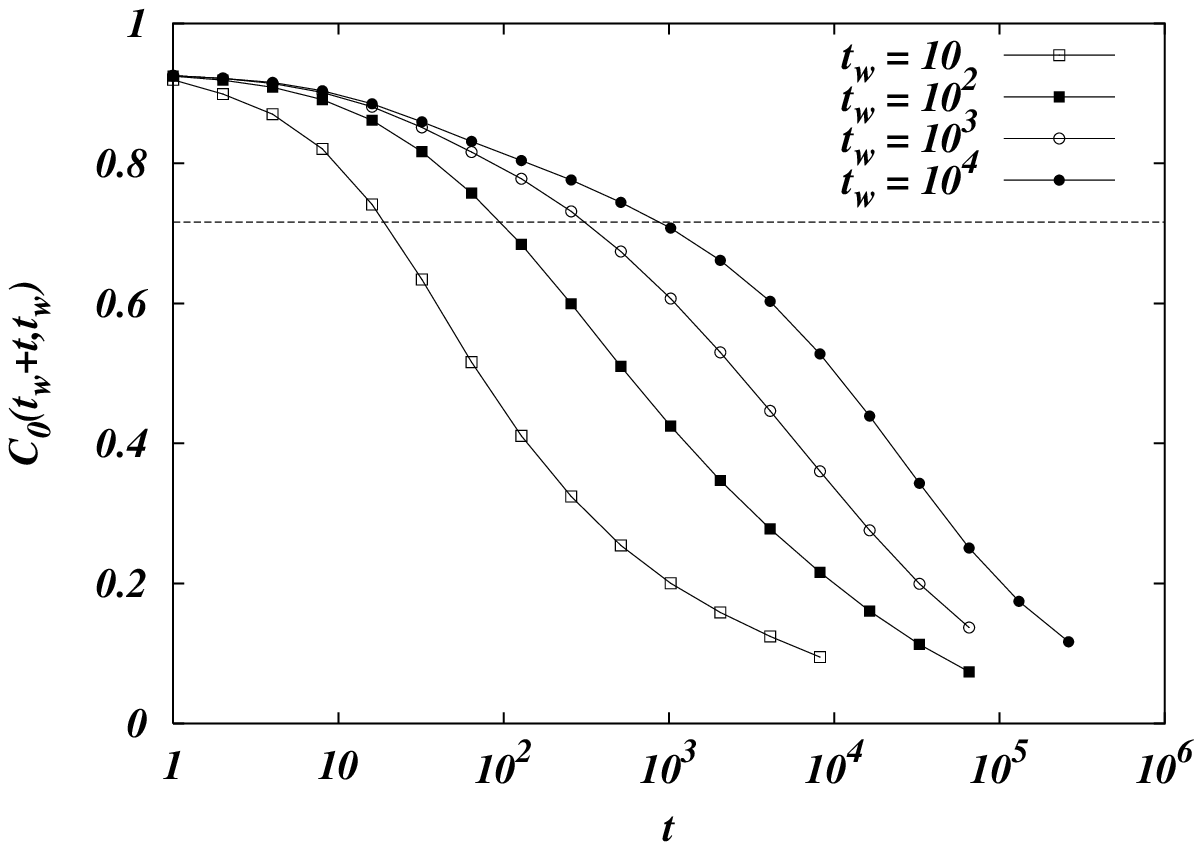}
\includegraphics[width=\columnwidth]{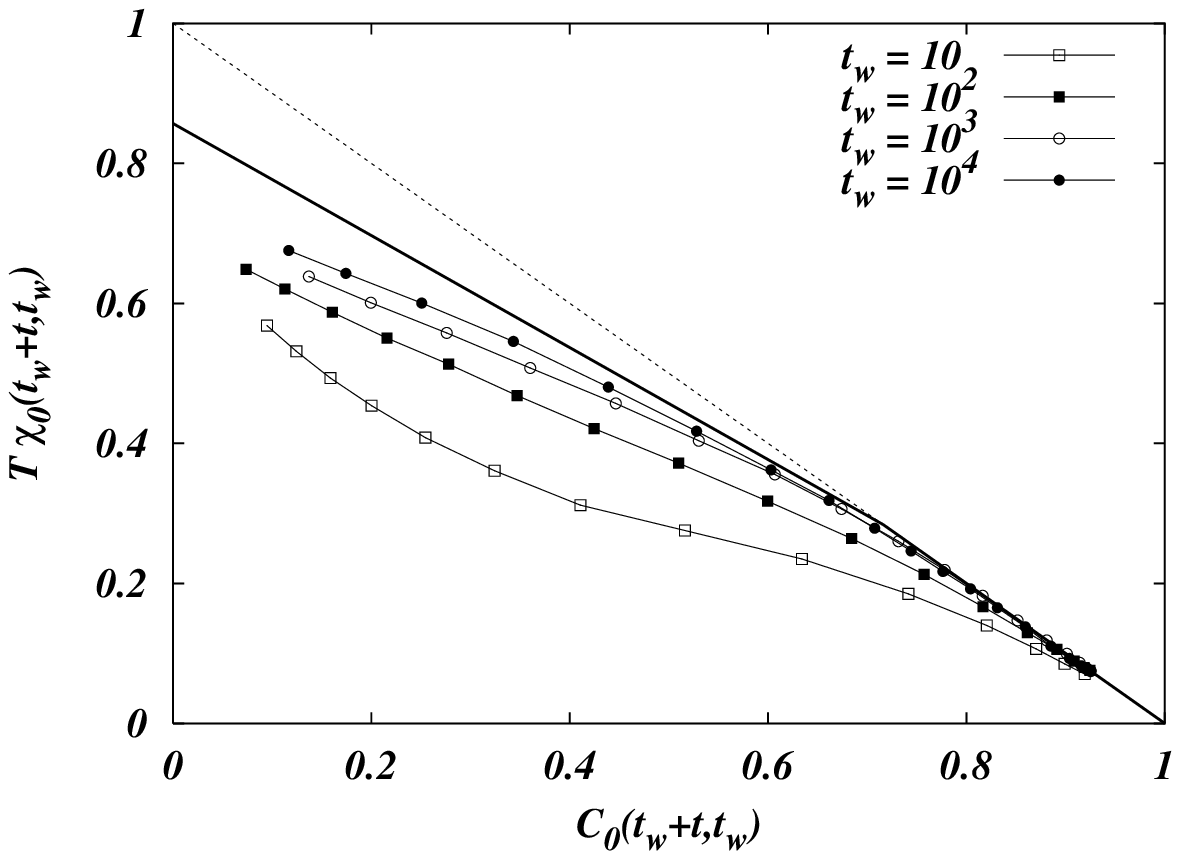}
\caption{The correlation function and the $\chi$-vs-$C$ plot for the
spin $i=0$ at $T=0.5$, which has $q_{\rm th}^{(0)}=0.716(7)$ (dashed
line, top).  Bold line (bottom) is the static $\chi_0[C]$ curve.}
\label{site0}
\end{figure}

Let us hereafter focus on type-II spins, that is on glassy degrees of
freedom.  In Fig.~\ref{site0} we show the correlation function for a
generic type-II spin ($i=0$ here) and the corresponding $\chi$-vs-$C$
plot.  For this spin we have $q_{\rm th}=0.716(7)$, shown with a
dashed line in Fig.~\ref{site0} (top).  In the limit of very large
times we expect (in analogy with \cite{OFDR}) the OFDR $\chi_i(t,t') =
\chi_i[C(t,t')]$ to hold. Moreover the function $\chi_i[C]$ should be
related to static quantities \cite{FMPP}.  Numerical data, for this
and for all the other spins, seem to converge to the static curve
$\chi_i[C]$. This is an evidence for a strong link between static and
dynamic observables even at the level of {\em single degrees of
freedom}.

\begin{figure}[!ht]
\includegraphics[width=0.9\columnwidth]{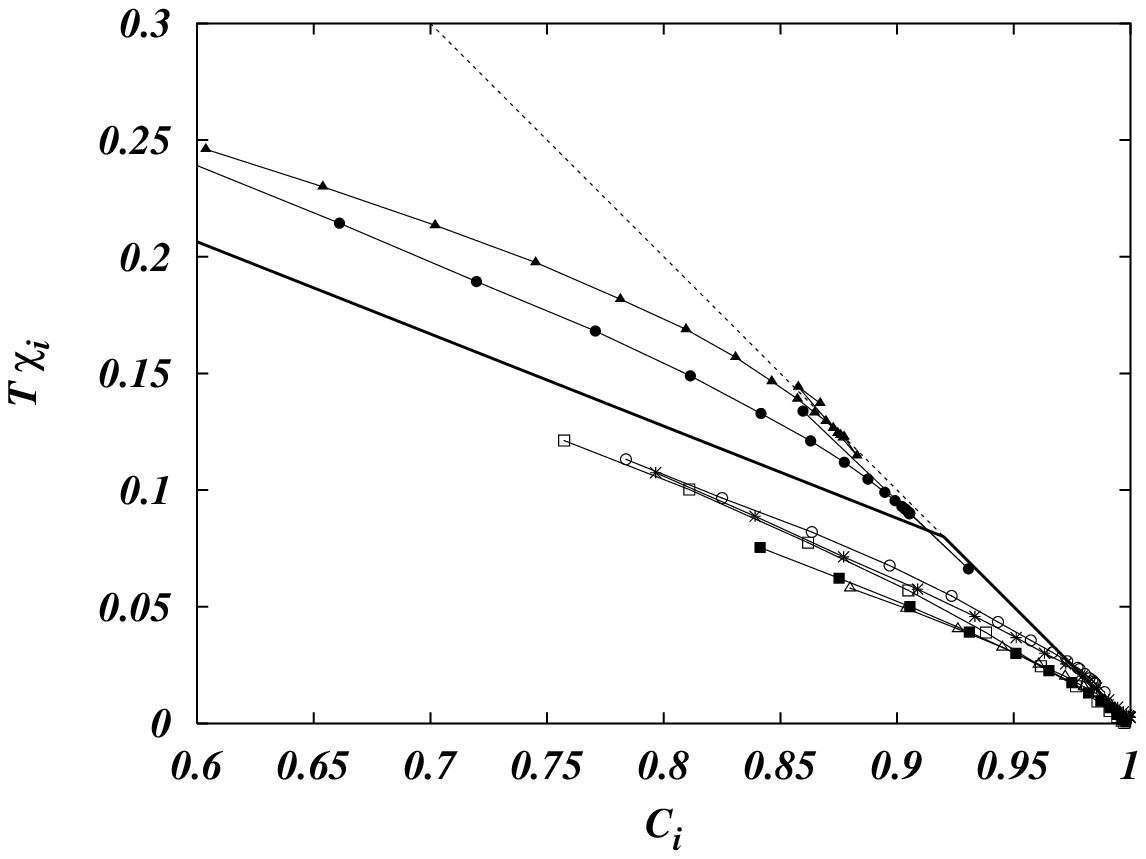}
\includegraphics[width=0.9\columnwidth]{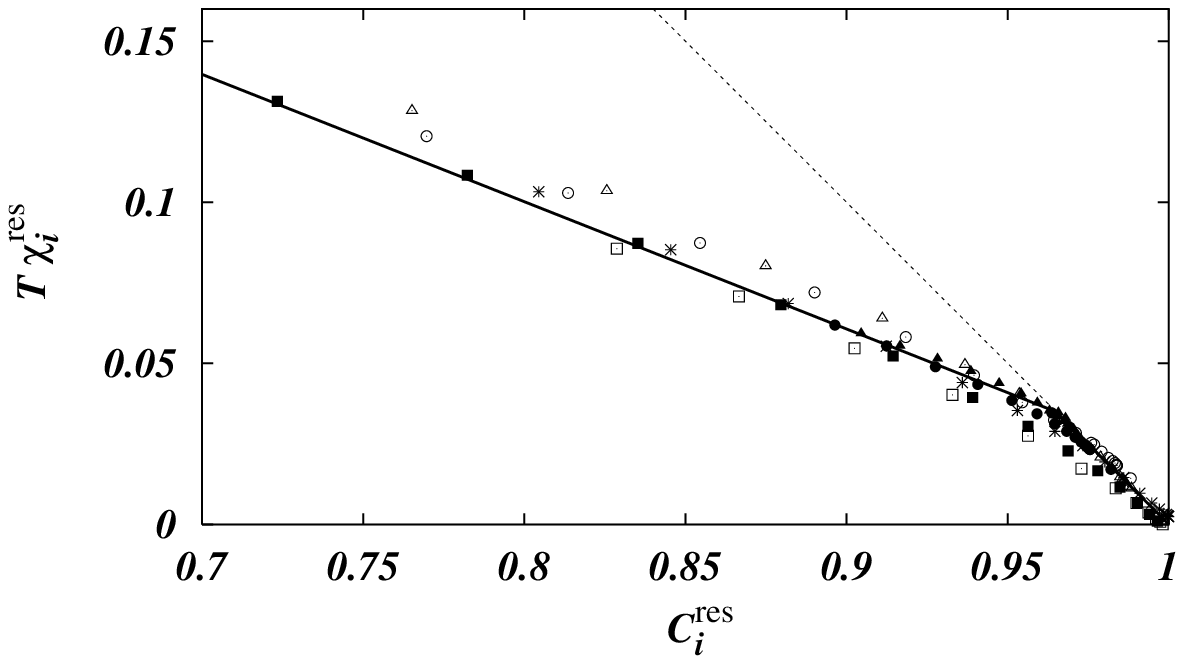}
\caption{In the $\chi$-vs-$C$ plot for seven generic sites at $T=0.3$
system heterogeneities become apparent.  The seven data sets can be
nicely collapsed with {\em no fitting parameters} (bottom).}
\label{7sites}
\end{figure}

The 1RSB static calculation yields
\begin{multline}
T\;\chi_i[C] = [1-C]\;\theta\left( C - q_{\rm th}^{(i)} \right) +\\
+ \left[1 - q_{\rm th}^{(i)} - m_{\rm th} \left(C - q_{\rm th}^{(i)}
\right) \right]\;\theta\left(q_{\rm th}^{(i)} -C \right)\, .
\label{eq:chi}
\end{multline}
The OFDR changes from site to site because the $q_{\rm th}^{(i)}$ 
changes.  Note, however, that the $\chi_i[C]$ curves are parallel in
the aging regime, since $m_{\rm th}$ only depends on the temperature
(cf. Fig.~\ref{7sites}).

System heterogeneities manifest themselves in the large variability of
the $q_{\rm th}^{(i)}$ local order parameters.  The $\chi$-vs-$C$ plot
for seven generic sites, see Fig.~\ref{7sites} (top), clearly shows
this variability.  In order to check that single-site data can be well
described by Eq.~(\ref{eq:chi}) we rescale data of Fig.~\ref{7sites}
using the scaling variables $C_i^{\rm res} = 1 - A_i (1-C_i)$ and
$\chi_i^{\rm res} = A_i \chi_i$, where $A_i = (1-\overline{q}) /
(1-q_{\rm th}^{(i)})$, $\overline{q}$ being a reference overlap which
can be chosen freely.  Rescaled data are shown in Fig.~\ref{7sites}
(bottom).

\begin{figure}[!ht]
\includegraphics[height=0.9\columnwidth,angle=-90]{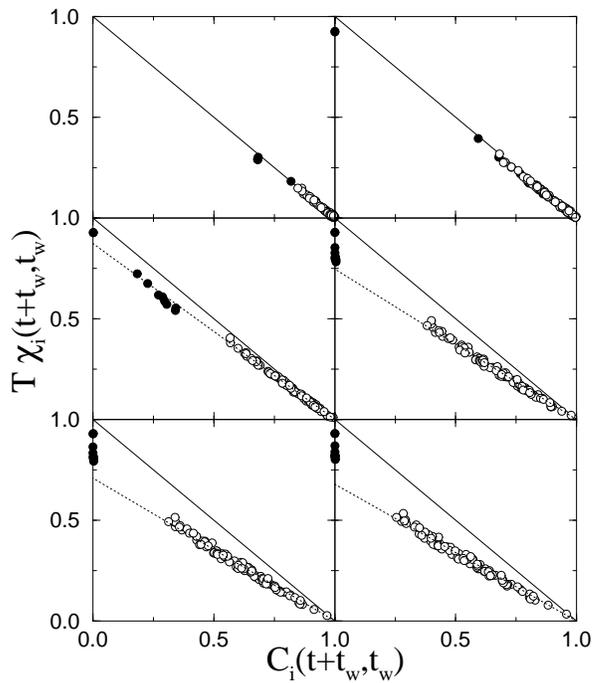}
\caption{The ``movie'' plot: Evolution of the single-spin correlation
and response functions in the $(C,\chi)$ plane. Here we use $h=0.1$,
$T=0.4$, and $t_w=10^5$. Different frames correspond to (from left to
right and up to bottom): $t=2^4$, $2^9$, $2^{12}$, $2^{15}$, $2^{16}$,
and $2^{17}$. Black and white circles refer, respectively, to type-I
and type-II sites. Black circles are not exactly on the FDT line
because of finite-$h_0$ effects, to be discussed in
Ref.~\cite{FDTSS2}. Dotted lines are fits to the white data.}
\label{movie}
\end{figure}

A last important question, in order to complete the description of the
time evolution of single-site quantities, regards the time law with
which spin $i$ runs along the curve $\chi_i[C]$.  The answer to this
question is very surprising and it is shown in Fig.~\ref{movie}, where
we plot the time evolution of all the 100 $C_i$'s and $\chi_i$'s.

Amazingly, sites-II data leave the FDT line {\em coherently} (when the
system undergoes a global structural rearrangement) and they remain
{\em very well aligned} for later times.  Fits to a function
\begin{equation}
\chi_i(t_w+t,t_w) = \frac{1-C_i(t_w+t,t_w)}{T_{\rm fit}(t_w,t)}
\end{equation}
give very accurate results with $T_{\rm fit} = 0.459$ (for
$t=2^{12}$), $0.536$ ($t=2^{15}$), $0.564$ ($t=2^{16}$), $0.590$
($t=2^{17}$).

In the following discussion we shall try to outline a few general
properties which can be extrapolated from our numerical results, and,
possibly, applied to a wider variety of systems.  Our basic objects
are the local correlation and response functions $C_i(t,t')$ and
$R_i(t,t')\equiv -\partial_{t'}\chi_i(t,t')$.  Following
Refs.~\cite{OFDR}, we guess that $\partial_t C_i(t,t'),\,\partial_t
R_i(t,t')\le 0$, and $\partial_{t'} C_i(t,t'),\,\partial_{t'}
R_i(t,t')\ge 0$. Moreover $C_i(t,t'),\, R_i(t,t')\!\to\!0$ as
$t\!\to\!\infty$ for any fixed $t'$. All these properties are well
realized within our model.

The {\bf first} non-trivial {\bf property}~\footnote{A somewhat
similar statement appears in L.F.~Cugliandolo, J.~Kurchan, and
P.~Le~Doussal, Phys. Rev. Lett. \textbf{76}, 2390 (1996).} is that,
for any ``non-exceptional'' pair of spins $i$ and $j$ there exist two
continuous functions $f_{ij}$ and $f_{ji}$ such that
\begin{equation}
C_i(t,t') = f_{ij}[C_j(t,t')]\ , \quad
C_j(t,t') = f_{ji}[C_i(t,t')]\ ,
\label{Equivalence}
\end{equation}
asymptotically for $t,t'\gg 1$.  We shall not specify what does
``non-exceptional'' mean~\footnote{One possibility is to take
Eqs.~(\ref{Equivalence}) as the definition of {\it dynamically
connected} degrees of freedom.  Our discussion is therefore valid
within each {\em dynamically connected component}.}, but the reader is
urged to bear in mind the example of type-I (paramagnetic) spins in
our model: if $i$ is type-I, and $j$ is type-II, then relation
(\ref{Equivalence}) clearly does not hold.

It is easy to show that transition functions $\{ f_{ij}\}$ can be
written in the form $f_{ij} =f_i^{-1}\circ f_j$.  Of course the
functions $f_i$ are not unique: in particular they can be modified by
a global reparametrization $f_i\to g\circ f_i$.

Moreover Eq.~(\ref{Equivalence}) implies a one-to-one correspondence
between the correlation scales (in the sense of \cite{OFDR}) of sites
$i$ and $j$.  Notice that, for our model, this is unavoidable if we
want the connection between statics and dynamics to be satisfied both
at the level of global and local (single-spin) observables.
Physically this first property means that structural rearrangements
occurs coherently in the whole system. Notice that this is not
unphysical, because far apart degrees of freedom are coherent only on
a coarse time resolution (diverging with $t_w$).

The {\bf second property} has been illustrated at length above: For
large times $t,t'$, a local OFDR of the form $\chi_i(t,t') =
\chi_i[C_i(t,t')]$ exists (the connection with the static result is
not a crucial point here).

Our {\bf third property} determines the form of the transition
functions $f_{ij}=f_i^{-1}\circ f_j$ in the aging regime.  In fact the
results in Fig.~\ref{movie} suggest that
\begin{equation}
\frac{\chi_i(t,t')}{1-C_i(t,t')} = \frac{\chi_j(t,t')}{1-C_j(t,t')}
\quad \forall\ i,j \ .
\label{NewTemperature}
\end{equation}
Combined with the OFDR, this means that we can take $f_i[C] =
\chi_i[C]/(1-C)$ for $C<q_{\rm th}^{(i)}$.  This relation cannot be
extended to the quasi-equilibrium regime $C_i>q_{\rm th}^{(i)}$,
because we would obtain $f_i[C] = \beta$ identically, which is not
invertible.

Finally the {\bf fourth property} is:
\begin{equation}
\chi_i'[C_i(t,t')] = \chi_j'[C_j(t,t')]\ ,
\end{equation}
or, in other words $\chi_i'[C_i] = \chi_j'[C_j]$ when
$C_i=f_{ij}[C_j]$.  Notice that, for our model, this property is
satisfied by the prediction we draw from the statics. However, a
direct numerical verification is quite hard.

Suppose now to measure the temperature of the spin $i$, by weakly
coupling it to a thermometer~\footnote{Here we consider a thermometer
based on the zeroth law of thermodynamics \cite{CUKUPE}.}. If the
thermometer respond quickly, it measures the bath temperature $T$ for
any site $i$, independently of the details of the thermometer
itself. However, if the thermometer is ``slow'', it measures an
effective temperature $T_{\rm meas}>T$, which depends upon the
specific thermometer used \cite{EXARTIER}.  Nevertheless, the above
scenario imply \cite{FDTSS2} that $T_{\rm meas}$ is site-independent.
The converse is also true: if one of the above four properties ceases
to hold, one can construct a thermometer which distinguishes colder
sites from hotter ones, and use it to transfer heat.

\begin{figure}[!ht]
\includegraphics[width=0.9\columnwidth]{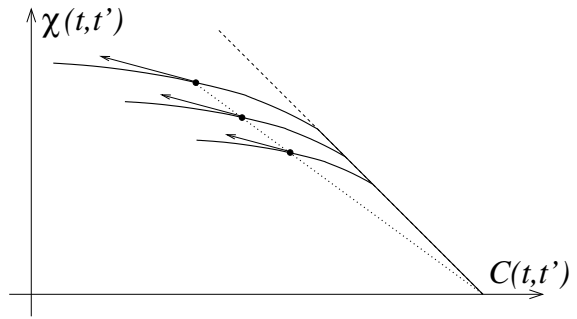}
\caption{A pictorial view of heterogeneous aging dynamics.  It is
sufficient to know the dynamics of a single spin in the system, in
order to reconstruct the behavior of any other one (once the static
parameters $q_{\rm th}^{(i)}$ are known).}
\label{ChiCorr}
\end{figure}

Our conclusions are summarized in Fig.~\ref{ChiCorr}.  Dynamical
heterogeneities turn out to be strongly constrained in the aging
regime.  These constraints can be derived from the hypothesis of
thermometric indistinguishability of different sites \cite{FDTSS2}.

We are deeply indebted with Riccardo Zecchina and Marc M\'ezard, who
discussed with us their results \cite{MEPAZE} before publication.
A.~M. thanks Leticia Cugliandolo and Jorge Kurchan for their interest
in this work, and the ESF for financial support.

\end{document}